\documentclass{svproc}
\usepackage{cite}
\usepackage{amsmath,amssymb,amsfonts}
\usepackage{graphicx}
\usepackage{textcomp}
\usepackage{xcolor}
\usepackage{hyperref}
\usepackage{caption}
\usepackage{subcaption}

\usepackage{tabularx} 

\def\BibTeX{{\rm B\kern-.05em{\sc i\kern-.025em b}\kern-.08em
    T\kern-.1667em\lower.7ex\hbox{E}\kern-.125emX}}

\begin{document}

\title{A Cloud-Edge Continuum Experimental Methodology applied to a 5G Core Study}
\titlerunning{A Cloud-Edge Compute Continuum Experimental Methodology}

\author{Samuel Rac\inst{1} \and Rajarshi Sanyal\inst{2} \and Mats Brorsson\inst{1}}
\authorrunning{Samuel Rac et al.}
\tocauthor{samuel Rac, Rajarshi Sanyal, and Mats Brorsson}
\institute{SnT, University of Luxembourg,\\
    \email{samuel.rac@uni.lu, mats.brorsson@uni.lu},\\
    \and
    Proximus Luxembourg S.A.,\\
    \email{rajarshi.sanyal@proximus.lu}}

\maketitle

\begin{abstract}

    There is an increasing interest in extending traditional cloud-native technologies, such as Kubernetes, outside the data center to build a continuum towards the edge and between. However, traditional resource orchestration algorithms do not work well in this case, and it is also difficult to test applications for a heterogeneous cloud infrastructure without actually building it. To address these challenges, we propose a new methodology to aid in deploying, testing, and analyzing the effects of microservice placement and scheduling in a heterogeneous Cloud environment. With this methodology, we can investigate any combination of deployment scenarios and monitor metrics in accordance with the placement of microservices in the cloud-edge continuum. Edge devices may be simulated, but as we use Kubernetes, any device which can be attached to a Kubernetes cluster could be used. In order to demonstrate our methodology, we have applied it to the problem of \textit{network function placement} of an open-source 5G core implementation.

\end{abstract}

\section{Introduction}

Cloud-native technologies, such as Kubernetes~\cite{Kubernetes_Brewer}, have significantly improved the way to allocate infrastructure resources to applications. For developers of distributed applications, deployment is greatly simplified as the individual components, typically embodied as Docker containers, are automatically mapped to nodes in the cluster that make up the infrastructure. This technology also has the potential to improve resource utilization and reduce over-provisioning, which is otherwise common. Overall it leads to shorter deployment times and reduced costs for infrastructure.

Edge and fog computing~\cite{laroui2021edge} have been introduced to enable the deployment of (parts of) applications closer to the end-user in order to lower end-to-end latency, reduce data sent over the network, or improve privacy by keeping the data local. While this has obvious benefits, it also introduces new challenges. A software component can no longer execute anywhere in the compute infrastructure as the Edge nodes typically require specific formats and explicit placement. While data center nodes display limited kinds of heterogeneity, edge nodes come in many different forms and architectures. We need to extend the cloud-native paradigm from the data center to the edge. From the developer's perspective, deploying an application taking advantage of the edge should be as easy as deploying it in a cloud data center. However, the best placement of software components is often not clear and extensive experimentation is needed, both for the placement and for finding the right system architecture.

Currently, there is no established methodology to test performance of cloud-native applications that span from the data-center to the edge. Currently used methods, see section~\ref{section:related_work}, use either simulation, meaning that real distributed applications cannot be tested, or they do not allow the testing of geographical distribution or heterogeneous architectures. We present a novel methodology to build testbeds for real distributed applications deployed in a cluster where nodes might be of different types and we model geographic distribution by controlling bandwidth, and latency between the nodes.

The methodology leverages the power of public cloud infrastructures and Kubernetes so that any application which can be deployed using Kubernetes can be used as a workload. We can simulate different geographical localities of subsets of nodes by controlling the latency and bandwidth available in communication links between nodes. Thanks to that, the application loading the system can be deployed unchanged from one experiment to another. With this methodology, we can avoid the tedious and time-consuming process of building large physical testbeds while the software development process can be kept the same as for a real environment and the experiments are easily reproducible.

We demonstrate this methodology with a study of the placement of 5G core network functions in either i) the central cloud, ii) at the network edge, or iii) in an intermediate local data center. The methodology is, however, general and can be used in any other setting involving the edge to data center continuum. One example is the deployment of a cloud multiplayer gaming system. Since the methodology leverages Kubernetes, it can run every containerized application. In that manner, the testbeds generated by this methodology are application agnostic.

Our main contributions are:

\begin{enumerate}
    \item a methodology to study the impacts of deploying applications in a heterogeneous cloud environment~\cite{HetBed} that i) allows for real distributed applications to be executed and ii) which does not need expensive physical infrastructure developed, and
    \item a performance analysis of a 5G core installation while studying three 5G use cases deployed using different system architectures on a testbed generated by our methodology.
\end{enumerate}

\section{Related work}
\label{section:related_work}

Goshi et al. describe a testbed that highlights Inter-NF dependencies~\cite{InvestigatingInterNFDependencies}.
Kube5G is a cloud-native 5G testbed designed to handle the whole 5G stack~\cite{Kube5G}.
COPA is an orchestration framework for networking running above the Kubernetes layer~\cite{COPA}.
However, these three testbeds (and the others referred to in their study) are not meant for the evaluation of placement and performance of the applications with respect to heterogeneous system architectures. It is not possible to simulate the impact of geographical distances between nodes on networking (e.g., latency) or the bandwidth restrictions. In contrast, our methodology enables the deployment of reproducible experiments in a public or private cloud without the costs and constraints of handling a country-sized network.

The \textit{AccessOpt} architecture detailed in section~\ref{architectures} is based on previous studies, e.g.~\cite{ElasticResourceManagement,  ANovelIoTArchitecture, CharacterizationAndIdentification, SDNFV5GIoT}. These studies describe multi-layered 5G architectures and are based on geographic areas and topologies as well as on logical layers. We do not claim to "invent" this architecture rather using a well-known architecture to demonstrate the capability of the methodology.

Sarrigiannis et al. describe a two-tier architecture (Cloud and Edge) for virtual NF placement with a VNF orchestrator~\cite{ApplicationandNetworkVNFmigration}. Contrary to their approach, we leverage Kubernetes, the state-of-the-art orchestration framework. It renders the flexibility to scale up and down on-demand or automatically. Exploiting Kubernetes, intricate architectures requiring complex interactions between nodes either at the control or user plane can be set up and tested without affecting the application.

Ejaz et al. present a three-tier architecture (Cloud IoT, Edge IoT, and Local Edge IoT) to improve reliability for mission-critical processes, based on \textit{iFogSim} simulator~\cite{MultilayerIoTEdgeArchitecture}. This study helped us to define our system architectures. However, the iFogSim simulator does not allow deploying a real containerized application.

Edgenet, as described by Şenel et al.~\cite{edgenet:2021}, provides a global distributed Kubernetes cluster, but it is not suitable as a testbed for 5G core or other edge-based applications as it cannot be configured, and there is no access to the Edge nodes.

Enoslib~\cite{enoslib:2021} is another suggestion to facilitate experimentation with distributed systems. It is a general tool to facilitate reproducible experimentation and is thus orthogonal to our methodology, which could be used as the backend in an Enoslib experiment. We have so far not seen it beneficial to use Enoslib.

\section{Methodology}\label{section:Methodology}

Our methodology relies on two main components: \textit{cloud-native technologies} and \textit{tools to simulate many architectural options} in a cloud environment. Testbeds according to this methodology can easily be deployed in public clouds. The tools and scripts needed for this are publicly available on github~\cite{HetBed}.

\subsection{Cloud-native technologies}

A testbed in our methodology is a distributed computer cluster that can simulate heterogeneous architectures and relies on well-known cloud-native technologies.

\textbf{Containers} We use Docker containers which greatly simplify application deployment~\cite{ahmed2018docker}. With a very lightweight virtualization layer, this technology has become a standard to package applications for deployment. To quickly deploy, scale up/down, and manage \textit{microservices} in a cloud environment, we use \emph{Kubernetes}, the state-of-the-art container orchestration tool, as mentioned earlier. Kubernetes manages \textit{pods} composed of at least one container.

\textbf{Monitoring} Cloud-native technologies contain a large set of tools for monitoring vast infrastructures. \emph{Prometheus} collects and exposes many metrics (CPU usage, memory, networking, and other metrics). Automatically, logs, network traces, and other metrics are effortlessly recorded and stored to be able to collect experimental metrics. In addition, a custom scheduler can use all the metrics collected to make better decisions.

\textbf{Kubernetes limitations} The Kubernetes scheduler is a powerful tool. It can find a proper \textit{microservices} placement when looking at available resources or node taints. However, network performance is not taken into account. It is not an issue while working within a traditional data center with homogeneous nodes, but it becomes a limitation when some nodes are outside the data center. It is, for instance, challenging to achieve ultra-low latency without considering at which geographical position a microservice is deployed.

\subsection{Architecture simulation}

In this section, we explain how we can simulate different system architectures on top of Kubernetes. This is a key feature for designing new infrastructures or developing new microservice placement strategies in the edge-cloud continuum.

\textbf{Node architecture} A crucial part of our methodology is the ability to run production-ready applications on top of the testbeds that we create. This means a testbed must consist of real compute nodes. These nodes should represent the nodes in the cloud-edge continuum we want to investigate. In our evaluation, we have been using a public cloud provider and are thus limited to the node types available at this provider. Currently, the choice of node types includes a range of ARM, Intel, and AMD processors with varying core counts. We can thus choose an ARM node with a small core count and (relatively) low amount of memory to represent an edge node, and larger Intel/AMD nodes can represent data center nodes. Obviously, this is not fully representing the range of possible architectures you might see in a real edge deployment, but for purposes of evaluating placement or scheduling options, this will be sufficient.

The nodes in the testbed are labeled according to their properties (resources, location, hardware accelerator) and follow a naming convention. These labels are used to select where to deploy an application's microservices according to system requirements using the Kubernetes scheduler.

\textbf{Configurability of Network Capacity and Latency} We also need to be able to represent the anticipated latencies and available network bandwidths in a geographically distributed cluster. Such configurable latency and bandwidth is a key feature of our methodology. It enables the simulation of distance and link capacity between nodes, e.g., between a data center and an edge node. Theoretically, the more the distance, the higher will be the latency.

The control of these parameters is achieved by means of \emph{traffic control}(\texttt{tc}). This is a utility program that can reconfigure the Linux kernel packet scheduler. It can add latency on received packets, change maximum bandwidth and other networking parameters. We run \texttt{tc} inside pods as a side-car, modifying the pod properties one by one.

\textbf{Microservice placement} The Kubernetes scheduler can use the above-described labels. Associating a microservice to a node can be done manually or automatically (implementing a custom scheduling policy).
Manual microservice placement is based on Kubernetes \textit{taints} and \textit{tolerations}, i.e., checking node labels and service permissions to know the candidate nodes where a service is authorized to be deployed. We can define a rule to force service deployment on a specific node using \textit{Pod affinity}.

\begin{figure}[t]
    \centering
    \begin{subfigure}[b]{0.48\columnwidth}
        \centering
        \includegraphics[width=\columnwidth]{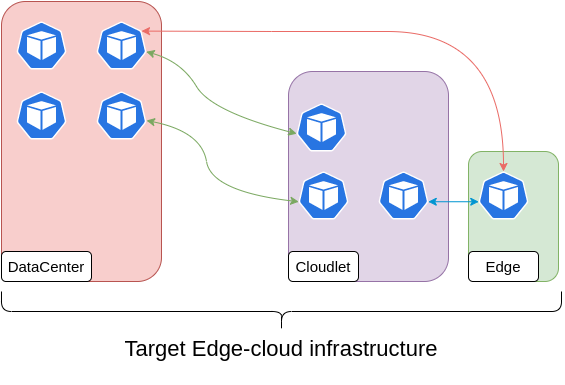}
    \end{subfigure}
    \hfill
    \begin{subfigure}[b]{0.48\columnwidth}
        \centering
        \includegraphics[width=\columnwidth]{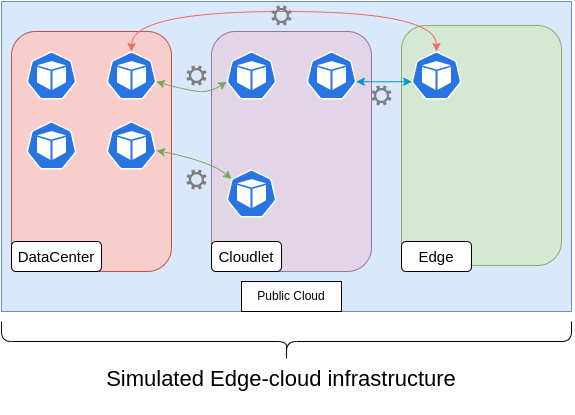}
    \end{subfigure}
    \hfill
    \caption{Edge-to-cloud environment can be simulated on the public cloud.}
    \label{HetBedarchitecture}
\end{figure}

Figure~\ref{HetBedarchitecture} gives an overview of the architecture simulation in a testbed setup. Nodes are labeled according to their kind, and networking between pods is configurable.

\section{Deploying a 5G system}
\label{section:deploy5g}

In order to demonstrate the usefulness of our methodology, we have used it to define a sequence of testbeds that can run edge computing experiments. The following sections describe how we conduct a study on a complete 5G core system implementation, studying the effects of 5G network function placement in different system architectures and for different use cases.

The $5^{th}$ generation (5G) of the cellular telecommunication network is amenable to being deployed in an edge-to-data-center continuum  (in contrast to previous generations, which needed much more specialized equipment). The main talked-about benefits of the 5G technology are enhanced Mobile Broadband (eMBB), Ultra-Reliable Low Latency Communication (URLLC), and massive Machine Type Communication (mMTC).

\subsection{5G Network Functions on the testbed}

A complete description of a 5G System (access network, devices, and core) is out of the scope of this paper, but some familiarity with the core components is necessary to understand the study. The 5G core consists of a number of \textit{Network Functions} (NFs). The gNodeB (gNB) represents the radio access network (RAN) to which user equipment (UE, e.g., phones) is connected over cellular radio. Most of the details about NFs are not important for this study, but we detail three of them: AMF, SMF, and UPF. These are essential to understanding how the system architectures are defined.

\begin{description}
    \item[AMF] (Access and Mobility management Functions) handles incoming connections and session requests of UEs and manages mobility (handover between two cells).
    \item[UPF]  (User Plane Function) handles user data traffic. The UPF is directly connected to a Data Network (Internet or Application Server).
    \item[SMF] (Session Management Function) establishes PDU sessions (Protocol Data Unit) for the UEs. A PDU session is a data tunnel that links a UE to a data network (DN) through a UPF.
\end{description}

\subsection{System architectures}\label{architectures}
In this study, we use three different kinds of nodes: \textit{Data center}, \textit{Edge} and \textit{Cloudlet} nodes, and define three system architectures comprising of different node types and topology. One architecture is used as a reference reproducing the traditional approach where all NFs are deployed in a data center, while the others use an edge node close to the gNB and a cloudlet node in-between the edge and the data center. We then experiment with different placement of the UPF, SMF, and AMF network functions in the different architectures and study the effect on system performance.

\begin{figure*}[t]
    \centering
    \begin{subfigure}[b]{0.3\textwidth}
        \centering
        \includegraphics[width=\textwidth]{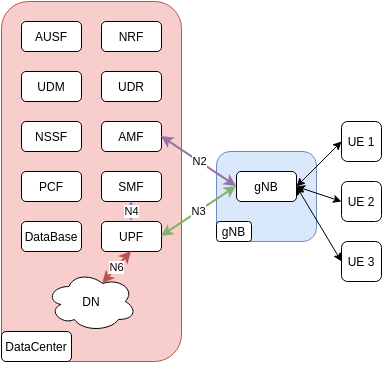}
        \caption{Baseline}
        \label{homogeneous}
    \end{subfigure}
    \hfill
    \begin{subfigure}[b]{0.3\textwidth}
        \centering
        \includegraphics[width=\textwidth]{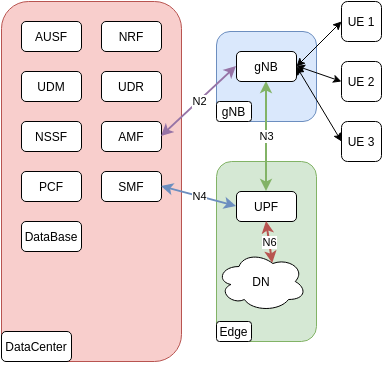}
        \caption{LatOpt}
        \label{ultra_low_latency}
    \end{subfigure}
    \hfill
    \begin{subfigure}[b]{0.3\textwidth}
        \centering
        \includegraphics[width=\textwidth]{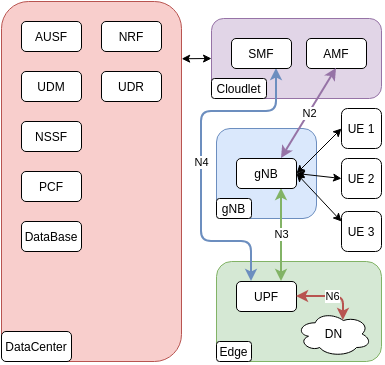}
        \caption{AccessOpt}
        \label{mIoT}
    \end{subfigure}
    \caption{Three different system architectures: a) Baseline, b) optimized for end-user latency and bandwidth, c) optimized for session throughput.}
    \label{Architectures}
\end{figure*}

Figure~\ref{Architectures} shows different architectures. For each architecture, the RAN elements (gNB and UEs) are deployed on separate nodes to not interfere with the NF placement study. Links between nodes are called N2 to N6 as defined in 5G standard architecture~\cite{3gpp.23.501}.

The \textbf{Baseline} architecture, shown in Figure~\ref{homogeneous}, is a reference architecture where all network functions are placed in the same data center. This architecture cannot support eMBB and URLLC use cases well (e.g., cloud gaming applications or AR/VR both need ultra-low latency and high bandwidth). The User Plane Function needs to be placed at the edge to achieve ultra-low latency.

The \textbf{LatOpt} system architecture, shown in Figure~\ref{ultra_low_latency},  is a well-known architecture. It should enable eMBB and URLLC use cases, significantly improving link N3 latency and throughput. With this architecture, the UPF is deployed on an edge node close to the gNB. Other NFs are running on data center nodes.

The \textbf{AccessOpt} system architecture is similar to LatOpt architecture but includes Cloudlet nodes. This architecture wants to be a simplified implementation of the multi-layered 5G architectures mentioned in section~\ref{section:related_work}. Investigating the effects of this architecture could provide valuable information for implementing more complex ones. Cloudlet nodes are closer to the data center nodes than edge nodes. Several gNBs may be connected to one Cloudlet node. AMF and SMF are deployed on Cloudlet nodes because they handle UE connection, session management, and mobility procedures. Thus, Cloudlet nodes can handle UE's massive mobility (many UEs moving from one gNB to another) while keeping reasonable latency with other NFs located in the data center.

Figure~\ref{mIoT} shows AccessOpt architecture. Deploying AMF and SMF on Cloudlet nodes should improve UE registration, mobility, and PDU session establishment procedures performances.

\subsection{Use cases}

In order to experiment with major 5G features (eMBB, URLLC, and mMTC), we introduce three use cases related to 5G. We investigate different NF placement, as discussed, on the above-described system architectures using these use cases: Augmented Reality (AR), Industrial IoT (IIoT, e.g., sensors in a smart factory), and Massive IoT (MIoT). Studying these three use cases will bring valuable knowledge for i) building new infrastructures including slices at the 5G edge, and ii) developing new scheduling methodologies for placing NFs.

The AR and IIoT use cases are detailed in-depth by Siriwardhana et al. in~\cite{PerformanceAnalysisofLocal5GOperatorArchitecturesforIndustrialInternet}. We adapt the workload and the experiment duration to the capabilities of the testbed. However, note that in our study, the 5G core and its NFs are not simulated but are real operation-grade elements.

In the AR use case, a UE should receive a high-quality video with low latency. We look at the UE end-to-end latency to evaluate different system architectures. The LatOpt architecture should improve this metric with respect to the baseline architecture by reducing the distance between the UPF and the gNB (manipulating latency).

For the Industrial IoT use case, we consider the UEs as sensors in a smart factory. In industry 4.0, we consider that an IIoT UE will not change of network cell and that the network is acquired at UE power up. Periodically, these devices will establish a data session and send their data to a processing server. Before sharing data, a session has to be established. Power constraints are not considered in this use case, the factory environment should provide energy to devices.
To evaluate the performance of this use case, we measure the end-to-end latency. The IIoT workload can be decomposed as follow: establishing a PDU session (to contact a processing server via the DN), sending data to the server, and getting the server's response. The IIoT end-to-end latency comprises two main parts: network acquisition time and data throughput (data transfer and server processing time). The AccessOpt architecture should have an impact on E2E latency for this use case. AMF and SMF located on a Cloudlet node should reduce the PDU session establishment time, while a UPF closer to gNB should reduce the data session's latency.

For the Massive IoT (MIoT) use case, we are evaluating the control plane's performances when connecting many UEs. These devices will generate traffic on the control plane when switching on/off (to save battery) or moving from one cell to another. In order to reduce the time to complete registration and session establishment procedures and to limit traffic toward datacenters, we deploy AMF and SMF on cloudlet nodes (according to AccessOpt architecture). Cloudlets should provide many benefits: i) being closer to the UEs than datacentres, ii) having more resources than edge nodes (to be able to scale up NFs if necessary), and iii) being close to many gNB at the same time to handle user mobility. Looking at the time to complete a procedure is an important KPI to assure QoS and avoid procedure time out.

\begin{table}[h]
    \tiny
    \caption{Use cases and their characteristics.}
    \label{use_case_matrix}
    \centering
    \begin{tabularx}{\columnwidth}{|>{\raggedright\arraybackslash}X|>{\centering\arraybackslash}X| >{\centering\arraybackslash}X| >{\centering\arraybackslash}X|}
        \hline
        \bf Use cases        & \textbf{Favoured System architecture}   & \textbf{Type of workload}                                                  & \textbf{KPIs}                              \\ \hline \hline
        AR (Smart Factory)   & Baseline or LatOpt                      & High data rate on the UP                                                   & E2E latency                                \\ \hline 
        IIoT (Smart Factory) & Baseline or AccessOpt                   & PDU session establishment process + Low data rate the UP                   & E2E latency                                \\ \hline
        MIoT (Massive IoT)   & Baseline or AccessOpt (+ load balancer) & Registration + PDU session establishment process + Low data rate on the UP & Time to register + establish a PDU session \\ \hline
    \end{tabularx}
\end{table}

\section{Experimental methodology}\label{section:experimental_methodology}

In this section, we outline the experimental setup and parameters of the experiments, such as additional latency and use case workload.

\subsection{Experimental setup}

To test all use cases, we run all the experiments in a public cloud environment. We use a self-managed Kubernetes cluster with one master node and seven worker nodes. All of these machines have 2 CPUs and 4 GB of RAM. On this cluster, we run the open-source 5G core free5G~\cite{free5gc}. Every Network Function (NF) runs inside its own pod. User Equipment (UE) and gNodeB are simulated using an open-source RAN simulator~\cite{ueransim}.

\subsection{Additional latency}

As described above, we can set an additional latency between two nodes to reflect the physical latency in the target system architecture.
Table~\ref{Experimental_parameters_latency} summarizes the additional latencies used in experiments. These are the additional latencies to what is already experienced in the physical cloud infrastructure.

\begin{table}[t]
    \tiny
    \caption{Additional latency used in different system architectures}
    \label{Experimental_parameters_latency}
    \centering
    \begin{tabularx}{\columnwidth}{|>{\raggedright\arraybackslash}X|c|>{\centering\arraybackslash}X|>{\centering\arraybackslash}X|>{\centering\arraybackslash}X|>{\centering\arraybackslash}X|>{\centering\arraybackslash}X|}
        \hline
        \textbf{System Architecture} & \textbf{N2 (ms)} & \textbf{N3 (ms)} & \textbf{N4 (ms)} & \textbf{N6 (ms)} & \textbf{DC-Cloudlet (ms)} \\ \hline \hline
        Baseline                     & 12.5             & 12.5             & 0                & 0                & 0                         \\
        LatOpt                       & 12.5             & 1                & 12.5             & 0                & 0                         \\
        AccessOpt                    & 3.5              & 1                & 3.5              & 0                & 9                         \\ \hline
    \end{tabularx}
\end{table}

\subsection{Workload parameters}

Table~\ref{Experimental_parameters_load} summarizes the use cases' workload parameters of the different experiments. The IIoT and MIoT use cases workload should mainly be managed by the Control Plane (respectively on SMF and AMF). In contrast, the User Plane (UPF) should support the AR use case workload.

\begin{table}[t]
    \tiny
    \caption{Use case Workloads.}
    \label{Experimental_parameters_load}
    \centering
    \begin{tabularx}{\columnwidth}{|>{\raggedright\arraybackslash}X|c|>{\centering\arraybackslash}X|}
        \hline
        \textbf{Use Case} & \textbf{Workload size}                              & \textbf{\#UEs} \\ \hline \hline
        AR                & 460 Mbit of video stream sent to the UE             & 3              \\ \hline
        IIoT              & PDU session establishment requests + 640 kB of data & 20             \\ \hline
        MIoT              & UE registration requests +                          & 50             \\
                          & PDU session establishment requests + 100 kB of data &                \\ \hline
    \end{tabularx}
\end{table}

\section{Results}
\label{section:results}
\begin{figure*}[t]
    \centering
    \begin{subfigure}[b]{0.45\textwidth}
        \centering
        \includegraphics[width=\textwidth]{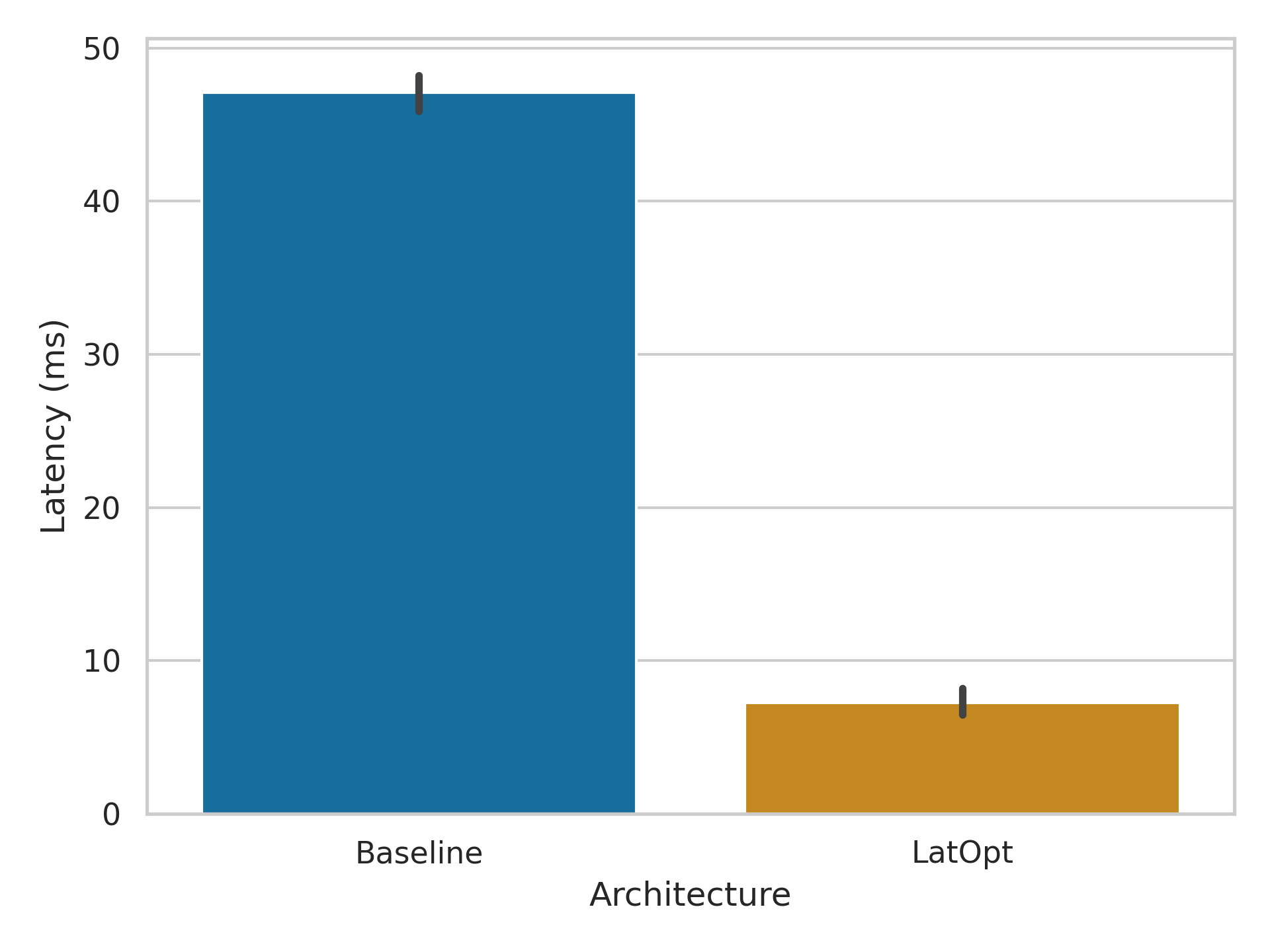}
        \caption{AR use case}
        \label{E2E_latency_AR}
    \end{subfigure}
    \hfill
    \begin{subfigure}[b]{0.45\textwidth}
        \centering
        \includegraphics[width=\textwidth]{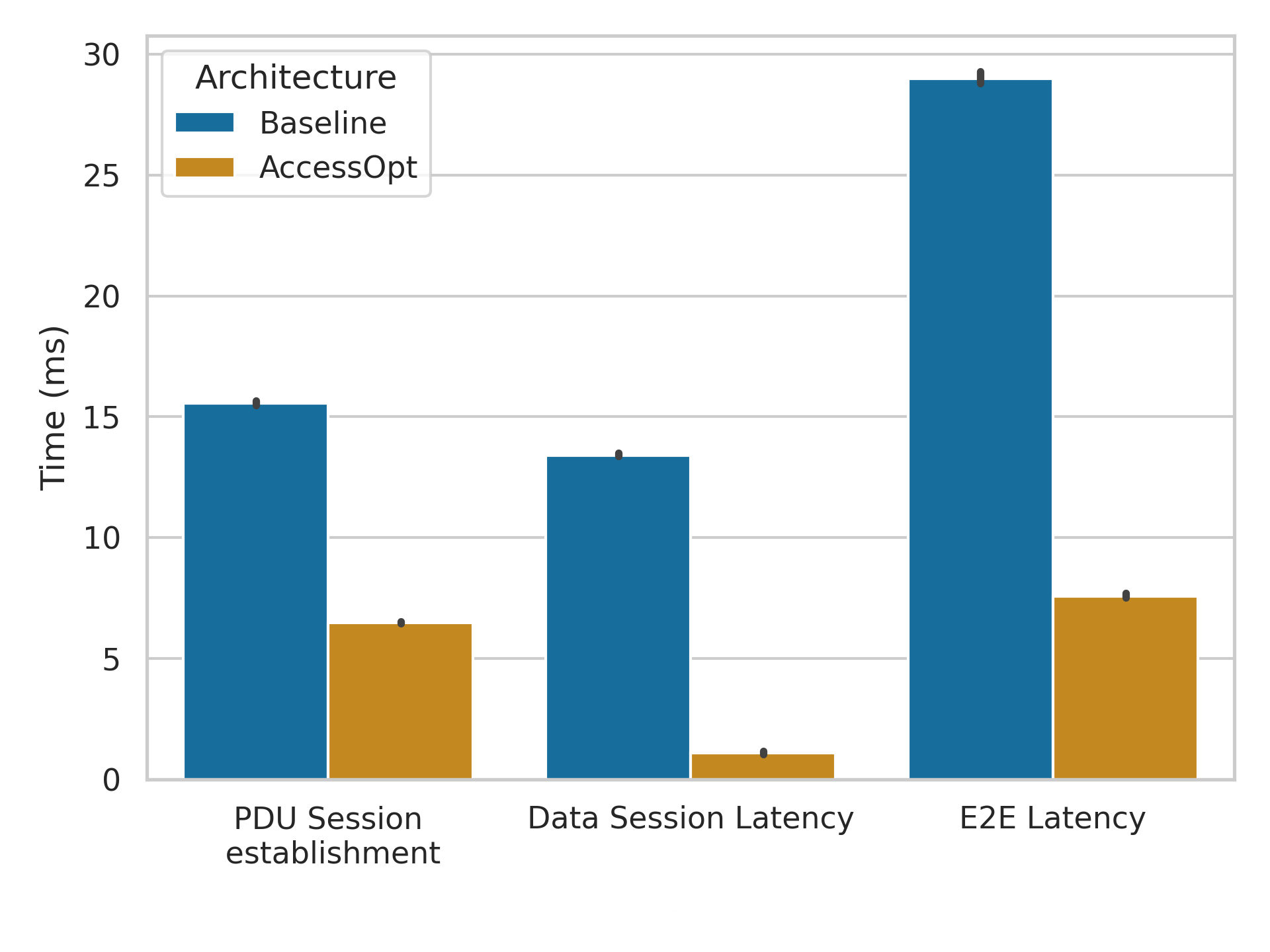}
        \caption{IIoT use case}
        \label{E2E_latency_IoT}
    \end{subfigure}
    \caption{End-to-end latency: a) AR use case, b) IIoT use case.}
    \label{E2E_latency}
\end{figure*}

\begin{figure}[t]
    \centering
    \includegraphics[width=0.45\columnwidth]{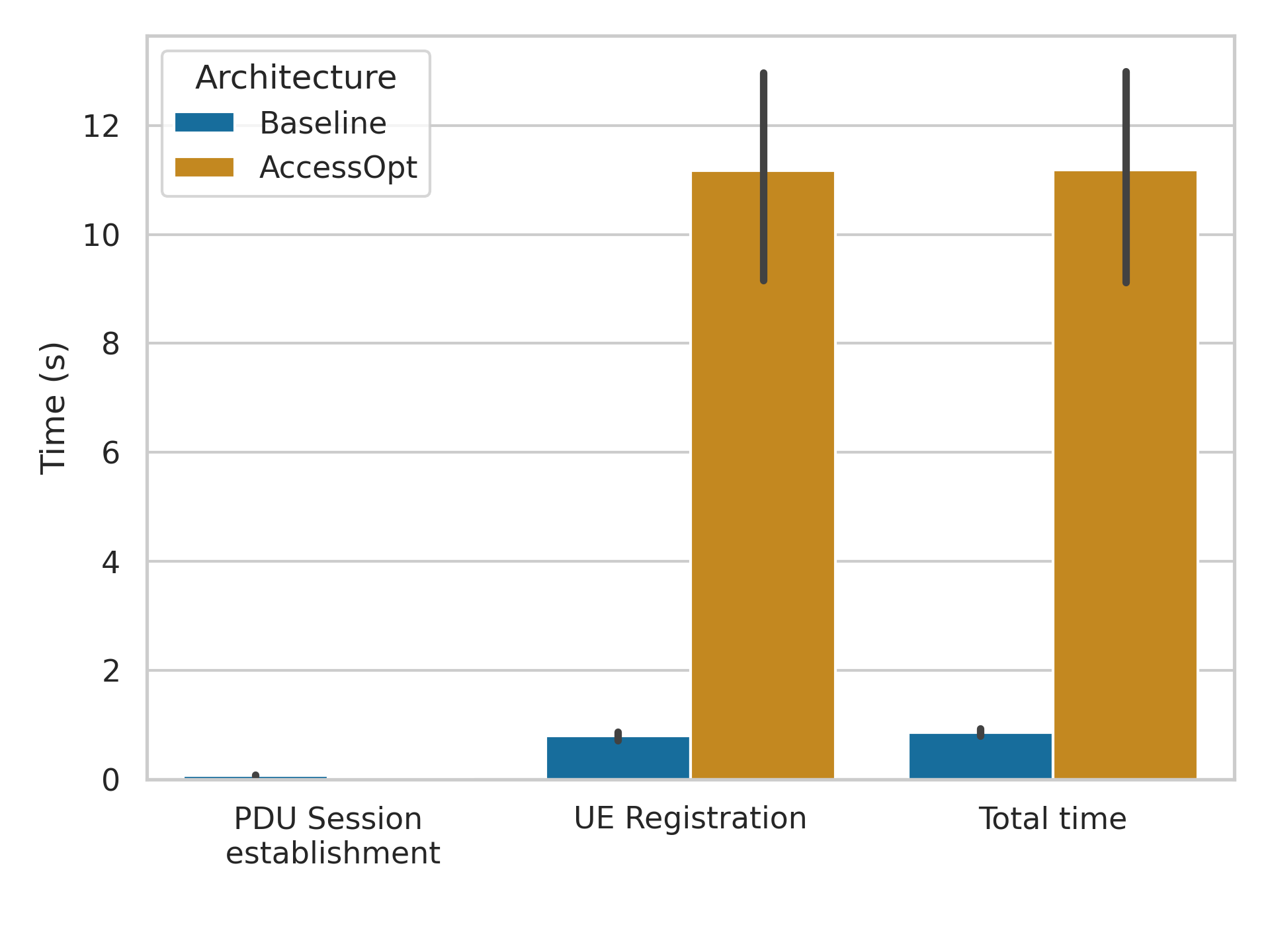}
    \caption{Duration of different procedures in the MIoT use case.}
    \label{procedures_duration_MIoT}
\end{figure}

In this section, we compare KPI values obtained using different architectures. These results provide insights into which architecture provides the best performance per use case. Figures~\ref{E2E_latency_AR},~\ref{E2E_latency_IoT} and~\ref{procedures_duration_MIoT} shows the mean KPI values for each use case according to the chosen architecture.

Figure~\ref{E2E_latency_AR} shows a significant difference in end-to-end latency for the AR use case. This KPI value is four times lower when using the LatOpt architecture. This improvement can be explained by positioning the UPF closer to the gNB. Latency on the link N3 is lower with the LatOpt architecture as well as end-to-end latency. This demonstrates the ability of the testbed by replicating well-known use cases.

The end-to-end latency for the IIoT use case is shown in figure~\ref{E2E_latency_IoT}. The AccessOpt architecture provides E2E latency almost four times lower than the Baseline architecture. E2E latency is divided into the time to achieve the PDU session establishment procedure and data traffic duration (transport and processing time). Both are significantly improved with AccessOpt. Establishing a new data session takes more time than transmitting data. Almost half of the requests go from AMF or SMF to the data center during the session establishment procedure, and the others go to the edge. Then having less latency to the edge improves the time to complete the procedure. With the AccessOpt architecture, the AMF is close to the edge nodes and on the same machine as SMF. This proximity explains the better KPI value for the AccessOpt architecture. Like in the AR use case, the UPF lowers the latency of the data session. The placement of AMF, SMF, and UPF in the AccessOpt architecture reduces the E2E latency significantly.

Figure~\ref{procedures_duration_MIoT} shows a significant difference in KPI values when using the Baseline and the AccessOpt architectures. The total procedure with the baseline architecture is 13 times faster than AccessOpt. UE registration procedure is 14 times faster on the Baseline architecture than on AccessOpt. However, the PDU session establishment procedure is ten times faster on the AccessOpt architectures. However, session establishment represents only 0.06\% of total time for AccessOpt and 8\% for Baseline. Therefore PDU session establishment time has a limited impact on AccessOpt architecture's total performance. Registration procedures have to be complete before a session can be established.
During the registration procedure, AMF mainly addresses NFs located in the data center (close to the database). This procedure will take more time to achieve with AccessOpt architecture, where AMF is far from the data center. It is contrary to the data session establishment procedure. Traffic is balanced between NFs in the Datacentre and at the Edge. Then, placing the AMF on a cloudlet node gives lesser performance for this procedure.

When the latency between the cloudlet and data center nodes becomes too high, it causes a systematic registration time-out. Only a few UEs can register before all registration timers are triggered when latency becomes high. In that case, UEs will try two more times to register without success. The UEs' procedure retries impact the CPU consumption of the control plane because a UE will initialize many procedures.

Our methodology helps to choose the best architecture for each 5G use case. Placing the UPF at the edge reduces the latency on the link N3 in every configuration tested. The optimal position of the AMF depends on the use case's procedures. AMF improves KPIs for the session establishment procedure when placed at the edge (or nearby), while results are better for the UE registration procedure when it stays in the data center.

\section{Conclusion}

Studying new scenarios in the edge-cloud continuum raises new experimental issues. Experimenters need testbeds that can reproduce every aspect of this heterogeneous environment. Our methodology aims to help deploy edge-cloud experience in a traditional cloud environment. We aim in the future to investigate custom Kubernetes schedulers, using this methodology to evaluate their performances.

\section*{Acknowledgments}

This work has been partly funded by the Luxembourg National Research Fund (FNR) under contract number 16327771 and has been supported by Proximus Luxembourg SA.

\bibliographystyle{spmpsci}
\bibliography{references}

\end{document}